\documentclass[lettersize,journal]{IEEEtran}
\usepackage{amsmath,amsfonts}
\usepackage{algorithmic}
\usepackage{algorithm}
\usepackage{array}
\usepackage[caption=false,font=normalsize,labelfont=sf,textfont=sf]{subfig}
\usepackage{textcomp}
\usepackage{stfloats}
\usepackage{url}
\usepackage{verbatim}
\usepackage{graphicx}
\usepackage{cite}
\usepackage{hyperref}
\usepackage{cite}
\usepackage{doi}
\usepackage{etoolbox}

\hypersetup{
    colorlinks=true,            
    linkcolor=blue,            
    citecolor=blue,            
    urlcolor=blue,              
    pdftitle={Your Title},      
    pdfauthor={Your Name},      
    pdfsubject={Your Subject},  
    pdfkeywords={Keyword1, Keyword2} 
}

\usepackage{mathtools}
\usepackage{xcolor}
\newcommand{\highlightgreen}[1]{%
  \colorbox{green!20}{$\displaystyle#1$}}

\begin{document}

\title{Toward quantum interconnects featuring nanometer-to-picometer bandwidth compression and THz-range quantum frequency conversion}

\author{Tim F. Weiss, Alberto Peruzzo
\thanks{Tim Weiss and Alberto Peruzzo are with the Quantum Photonics Laboratory and Centre for Quantum Computation and Communication Technology, RMIT University, Melbourne, VIC 3000, Australia}
\thanks{Alberto Peruzzo is with Quandela, Massy, France}}

\markboth{Manuscript daft V2, January~2025}%
{Weiss \MakeLowercase{\textit{et al.}}: Toward quantum interconnects featuring nanometer-to-picometer bandwidth compression and THz-range quantum frequency conversion}

\IEEEpubid{}

\maketitle

\begin{abstract}
The long-range transmission of quantum information relies on multiple interfaces between photons, acting as flying qubits, and localized memories, serving as repeaters, to mitigate transmission losses. Efficient, long-range transmission necessitates the use of short, picosecond-scale photons, which are markedly different from the narrowband, nanosecond-scale photons optimal for absorption by memory elements, typically operating at wavelengths far from telecom. In this article, we point toward designs capable of bridging these regimes, leveraging the interplay between sum-frequency generation-based quantum frequency conversion and resonant confinement in an integrated ring resonator. 
\end{abstract}


\section{Introduction}
The realization of quantum technologies has been pursued across several fundamentally different platforms, including optical architectures \cite{YinLi:20,ZhongPan:20,WangThompson:20}, solid-state systems \cite{ZwanenburgEriksson:13,WolfowiczAwschalom:21}, and superconducting circuits \cite{KjaergaardOliver:20}. Each of these platforms is uniquely suited for specific applications. While these platforms by themselves represent well-developed fields of research, their interconnection, which is crucial for the eventual realization of extensive quantum networks, has received considerably less attention.

The long-range transmission of quantum information, required for distributed quantum computing \cite{CuomoCacciapuoti:20}, quantum key distribution \cite{XuPan:20}, and quantum secure direct communication \cite{LongLiu:02}, is expected to require multiple interfaces between photons acting as flying qubits and localized memories to mitigate transmission losses \cite{AzumaTzitrin:23}. Reconciling the disparity between the spectral-temporal properties of photons ideal for long-distance transmission with those optimal for absorption by a memory or repeater element represents a key challenge of such interfaces.

Quantum memories based on single atoms \cite{ReisererRempe:15}, atomic ensembles \cite{HammererPolzik:10, ParigiLaurat:15}, trapped ions \cite{BruzewiczSage:19}, and solid-state defects \cite{AwschalomZhou:18} operate at wavelengths significantly different from the telecommunication bands. They involve photons with spectral-temporal properties determined by the excited state’s lifetime, typically in the nanosecond range. In contrast, photons acting as flying qubits need to be temporally short, in the picosecond range, to enable high-density packing for efficient transmission of information \cite{EssiambreGoebel:10,EsmaeilZwiller:21}.

Therefore, large-scale quantum networks will require interconnects between these two regimes, achieving both frequency shifts of hundreds of nanometers and compression of the spectral bandwidth down to the MHz range ($ \sim 0.1 \mathrm{pm}$), ideally in a single device. With telecom-to-memory frequency conversion already featured across a number of milestone experiments \cite{MaringRiedmatten:17,VanLeentWinfurter:22,KnautLukin:24}, simultaneous significant bandwidth compression, in particular, remains an unresolved issue.

While spectral bandwidth compression of up to three orders of magnitude has been demonstrated using electro-optic manipulation \cite{SosnickiKarpinski:23}, significant (quantum) frequency conversion \cite{Kumar:90} is achievable only through nonlinear interaction. Approaches attempting to achieve spectral compression alongside substantial frequency shifts have so far been limited to a single order of magnitude \cite{LavoieResch:13,AllgaierSilberhorn:17}.

In the following, we point toward a device that simultaneously achieves quantum frequency conversion (QFC) over hundreds of nanometers and spectral bandwidth compression by up to three orders of magnitude. This is accomplished through single-photon sum-frequency generation in integrated photonics circuits. We in particular build on works in the so far disconnected fields of narrowband photon generation and QFC, connecting both fields by adapting the former to single photon operation and presenting modeling and illustration to describe the operation of the latter under resonant confinement. 

With this paper, we aim to illustrate the theoretical concepts that would give the functionality of such a device, and to provide the general design required for experimental implementation, together with an analysis of essential design-considerations.

\section{Design}
\begin{figure*}
    \centering
    \includegraphics[width=2\columnwidth]{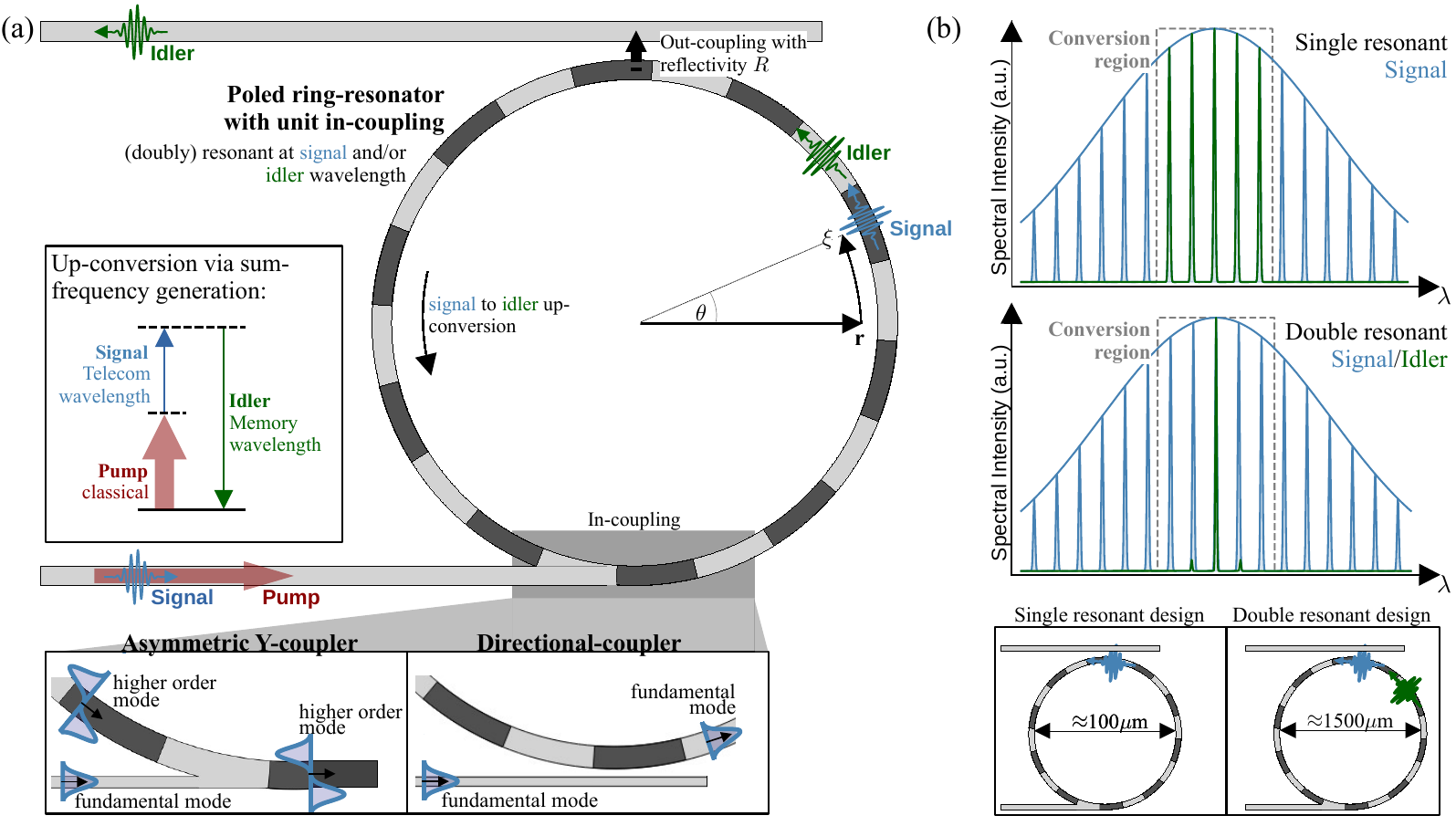}
    \caption{\textbf{Schematic of proposed devices.} (\textbf{a}) Device design featuring a ring resonator with an input waveguide configured so that the in-coupling of the signal photon can be achieved with near-unit probability, using either an asymmetric Y-coupler or a critically coupled directional coupler. The ring is poled to quasi-phase-match a quantum frequency conversion process between a broadband telecom signal photon and a fundamental mode idler photon at a memory-compatible wavelength. We note, that we have refrained from specifying explicitly device parameters like the waveguide geometry, which, while ultimately required for experimental implementation, are not essential to the device's functionality, and may be realized in a number of different ways. (\textbf{b}) Schematic illustration of the interplay between the resonances of the ring resonator and the generation of narrowband idler photons: for a single-resonant design, idler photons are generated at the resonances of the signal photon, while for the double resonant design efficient conversion occurs only if a set of signal and idler resonances happen to satisfy conservation of energy. The actual device is required to possess a conversion region larger than the bandwidth of the signal photon, while containing only a single, narrow resonance at the idler frequency.}
    \label{fig:DesignSketch}
\end{figure*}

The proposed device consists of a ring resonator with an input waveguide configured so that the in-coupling of the signal photon may be achieved, in principle, with near-unit probability. Within the resonator, QFC based on sum-frequency generation is mediated by a domain-engineered material nonlinearity. This, in combination with specifically designed interference effects resulting from confinement in the resonator, results in the generation of an idler photon with a narrow frequency bandwidth, as sketched schematically in Fig. \ref{fig:DesignSketch}. 

To achieve the near-unit in-coupling critical for devices operating on the single-photon level, the device features an asymmetric, multi-mode Y-coupler that populates the ring resonator with higher-order signal and pump modes \cite{ChenYang:14,SunZhang:17,LuZhang:19,ChangZhang:18,ZhangFu:17,LoveRiesen:11}. Alternatively, a more traditional critically coupled directional coupler may be used, but it is expected to lose parts of the signal wavepacket until the loss and gain mechanisms in the ring reach equilibrium.

In the following, we discuss in detail the operating principles of the device and the associated requirements. In particular, we separate the working mechanism into two parts: the quantum mechanical frequency conversion process mediated by material nonlinearity, described in Section II.A, and the formation of resonant modes within the ring resonator—a linear, classical effect—discussed in Section II.B. We note, however, that in the actual device both QFC and bandwidth compression occur simultaneously, within the same ring resonator.

\subsection{Quantum frequency conversion}
QFC based on three-wave-mixing processes presents a well-establish technology for the transduction of single photons from one frequency band to another without introducing uncontrolled changes to other state properties, and while preserving entanglement and photon statistics \cite{Kumar:90}. The nonlinear sum-frequency process between the pump, signal and idler modes with frequencies $\omega_{p,s,i}$ and $\omega_{p} + \omega_{s} = \omega_{i}$ considered here may efficiently be described by the Hamiltonian \cite{QuesadaSipe:22,ChristSilberhorn:13}
    \begin{equation}
    \label{eq:Hamiltonian}
    \begin{aligned}
        \hat{H}_{\mathrm{SFG}} \; = \; &- \hbar \eta \: \underbrace{ \highlightgreen{ \int \mathrm{d}\xi \: s_{\mathrm{nl}}(\xi) \mathrm{e}^{i (\beta_{p} + \beta_{s} - \beta_{i}) \xi} } \; \alpha_{p} (\omega_{p}) }_{\text{Process transfer function}} \\
        & \times \hat{a}_{s}(\omega_{s}) \hat{a}_{i}^{\dagger}(\omega_{i}) \; \; + \; \; H.c. ,
    \end{aligned}    
    \end{equation}
wherein the single-photon signal and idler fields are quantized and represented via photon annihilation and creation operators $\hat{a}_{s}$, $\hat{a}_{i}^{\dagger}$ respectively, corresponding to the discrete frequency modes of the ring-resonator; $\eta$ represents the effective coupling constant \cite{QuesadaSipe:20,QuesadaSipe:22}. We note, that this Hamiltonian describes the nonlinear QFC process only, and does not account for linear propagation within the resonator, which we consider separately in the section below. To occur efficiently, the QFC process requires a strong classical pump field.

\begin{figure*}
    \centering
    \includegraphics[width=2\columnwidth]{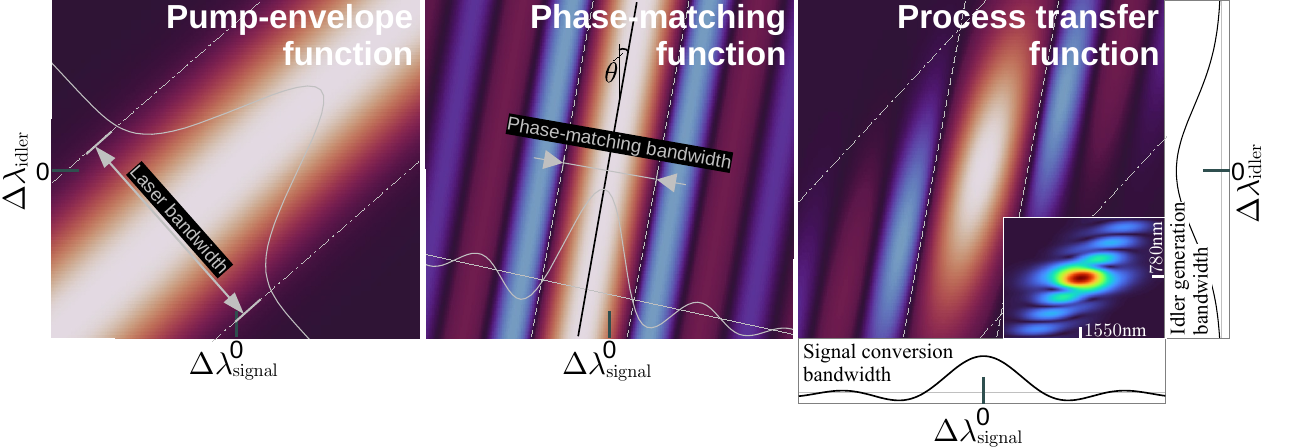}
    \caption{\textbf{Frequency conversion process.} Schematic illustration of how the characteristics of the QFC process arise from the device parameters in the Hamiltonian (\ref{eq:Hamiltonian}) and how they shape the conversion process: The pump-envelope function, determined by the spectral distribution of the pump $\alpha_{p}(\omega_{p})$, together with the phase-matching function, form the process transfer function (PTF), which fully characterizes the spectral properties of the sum-frequency conversion process. The width of the pump-envelope function is defined by the bandwidth of the corresponding laser, which takes on significant values only when a pulsed source is used. The width of the phase-matching function scales inversely with the (effective) length of the interaction region and can be broadened by applying a chirp to the periodic poling used for phase matching. The angle $\theta$ of the phase-matching function is determined by the group velocities of the interacting fields and thus depends on the wavelengths of the interaction fields and the device’s dispersion. To achieve a conversion bandwidth that covers the entire spectrum of the incident signal photon (approximately $5\,\mathrm{nm}$), it is essential to design the PTF by appropriately considering the width of the pump-envelope function as well as the angle and width of the phase-matching function. The inset in the right panel depicts a PTF calculated along theory developed by Quesada \textit{et al.} \cite{QuesadaSipe:20,QuesadaSipe:22}, adapted here for QFC \cite{ChristSilberhorn:13}, for an explicit waveguide with approximately asymmetric group-velocity matching ($\theta \approx 90^{\circ}$, conversion of a 1550$\,$nm signal to a 780$\,$nm idler photon, see Appendix A), including effects of time ordering, self- and cross-phase modulation associated with high(er) gain operation. Once these effects become significant, the PTF will show deviations from the product of the pump-envelope and the phase-matching function but will retain general features like bandwidths and orientation.}
    \label{fig:FrequencyConversion}
\end{figure*}

Due to the conservation of energy, the frequency at which the idler photon is generated is fully determined by the pump and the signal fields. Efficient operation further requires conservation of momentum, mediated by the function highlighted in green, here expressed via the waveguide propagation constants $\beta_{k}$. This term, together with the spectral distribution of the pump $\alpha_{p}(\omega_{p})$ allows to calculate the process transfer function (PTF), which fully describes the spectral properties of the conversion process. This gives the effective conversion bandwidth of the incident signal and the generation spectrum of the idler. To accommodate the entire bandwidth of the signal photon, the PTF needs to be designed appropriately, which requires either a broadband pump laser or an appropriately chirped poling pattern $s_{\mathrm{nl}}(z)$ \cite{Harris:07}. These requirements are illustrated in more detail in Fig. \ref{fig:FrequencyConversion}.

The efficiency of the sum-frequency generation process depends directly on the pump power and the effective path length of the resonator. This can be seen by considering the interaction mediated by the Hamiltonian (\ref{eq:Hamiltonian}) in the Heisenberg picture for an ideal, pure state \cite{IkutaImoto:11,IkutaImoto:13,ChristSilberhorn:13}
\begin{equation}
    \label{eq:Efficiency}
    \begin{aligned}
        \hat{A}_{s}^{\mathrm{(out)}} \; &= \; \mathrm{cos} \left( \eta \tau \right) \hat{A}_{s}^{\mathrm{(in)}} \; - \; e^{i\phi} \mathrm{sin} \left( \eta \tau \right) \hat{A}_{i}^{\mathrm{(in)}} \\ 
        \hat{A}_{i}^{\mathrm{(out)}} \; &= \; e^{-i\phi} \mathrm{sin} \left( \eta \tau \right) \hat{A}_{s}^{\mathrm{(in)}} \; + \; \mathrm{cos} \left( \eta \tau \right) \hat{A}_{i}^{\mathrm{(in)}},
    \end{aligned}
\end{equation}
wherein $\hat{A}_{s/i}^{\mathrm{(in/out)}}$ represent annihilation operators of the respective (broadband) photon modes, $\phi$ the phase of the pump, $\eta$ the effective interaction strength proportional to the pump power, and $\tau$ to the nonlinear interaction time, proportional to the effective path length. Accordingly, the probability of converting the input photon from the signal to the idler mode is given by $\mathrm{sin}^{2} \left( \eta \tau \right)$, which, upon appropriate selection of the pump power, allows to tune the process to occur with near unit efficiency, as has been demonstrated experimentally \cite{BockEschner:18}.  Due to the long effective interaction length resulting from confinement in the resonator, this should be possible for relatively low pump powers, reducing undesired time-ordering effects associated with high-gain operation.
We note, that the theoretical conversion efficiencies calculated from (\ref{eq:Efficiency}) often differ significantly from those obtained experimentally due to non-uniformity of the fabricated waveguides \cite{ChenFan:24}, which, in an experimental realization, would have to be accounted for by further adjusting the pump power. The QFC conversion process may further be modified by self- and cross-phase modulation induced by a strong pump field \cite{QuesadaSipe:22}, and depends intricately on the overlap of pump and signal temporal modes \cite{BrechtSilberhorn:11, QuesadaSipe:16}.

\begin{figure*}
    \centering
    \includegraphics[width=2\columnwidth]{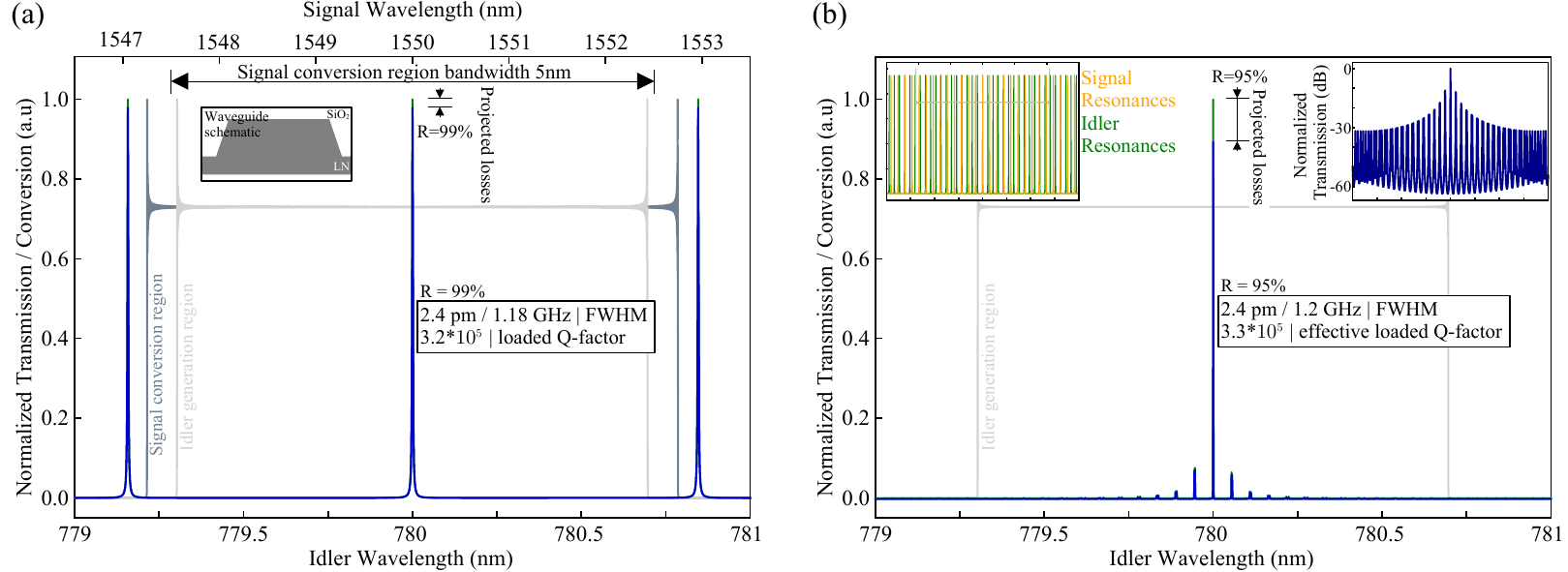}
    \caption{\textbf{Frequency compression process.} 
    (\textbf{a}) Resonant response of the singly-resonant design featuring a small resonator with radius $r \approx 51\,\mathrm{\mu m}$. 
    (\textbf{b}) Effective resonant response of a large resonator with a radius an order of magnitude larger than the design in (a), doubly resonant at both the desired signal and idler frequencies ($r \approx 741\,\mathrm{\mu m}$). The inset at the top-left depicts the full resonant response of the device, while the inset at the top-right depicts the device's response considering, additionally, the conversion process enforcing conservation of energy, as shown in the main plot, on a logarithmic scale. We note that this implementation features undesired side-resonances. The signal-photon-conversion and idler-photon-generation regions corresponding to a slightly chirped poling period are indicated in gray in both figures, achieving uniform conversion over the full bandwidth of picosecond-scale signal photons. For both designs, waveguide dispersion based on state-of-the-art LNOI microring fabrication \cite{LuTang:19} is included (see Appendix A). Full-width-half-maximum (FWHM) idler photon bandwidths and loaded Q-factors calculated at the up-converted idler wavelength corresponding to the central resonance are indicated for different reflectivities $R$ at the output waveguide. Both figures include the projected impact of propagation losses due to scattering and transmission into the output waveguide, based on state-of-the-art LNOI waveguide fabrication \cite{ZhangLoncar:17}, calculated from the transfer functions of the respective devices. Both designs consider an exemplary frequency conversion process between $1550\,\mathrm{nm}$ telecom photons and $780\,\mathrm{nm}$ photons suitable for memories based on $\mathrm{\prescript{87}{}{Rb}}$ \cite{VanLeentWinfurter:22}.}
\label{fig:FrequencyCompression}
\end{figure*}

\subsection{Bandwidth compression}
Converting the incident telecom photon from the signal frequency to the idler frequency with significant bandwidth compression requires two primary conditions:

(\textit{1}) The resonances associated with the QFC process need to be spaced far apart, so that only one resides within the conversion region of the nonlinear interaction. Here, we consider frequency up-conversion over a range of $5 \mathrm{nm}$, able to accommodate a transform-limited broadband telecom photon with a temporal length on the order of a single picosecond.

(\textit{2}) The width of this resonance needs to be sufficiently narrow to qualify for absorption by memory elements. The resonance width is directly related to the number of round trips the photon undergoes before leaving the resonator, which, in turn, depends on the coupling to the corresponding output waveguide and the propagation losses. Compression to spectral bandwidths on the order of $1 \mathrm{pm}$ would require resonators with loaded quality factors on the order of $10^{5}$ \cite{LuTang:19,ZhangLoncar:17}. 

The loaded quality factor of the resonator—and thus the degree of bandwidth compression—can be increased by reducing the coupling to the output waveguide (or vice versa), with the ultimate limit set by propagation losses associated with the increased effective path length. More generally, adjusting the loaded quality factor by manipulating propagation losses or the coupling to the output waveguide introduces a trade-off: as bandwidth compression increases, the probability of retaining the converted photon decreases.

In the following, we present two alternative designs based on a small single-resonant device and a larger double-resonant device. Both designs implement the same conversion process but rely on different operating mechanisms (see Fig. 1b). As a result, the small resonator requires only the signal to be resonant, which simplifies the design but, due to its small footprint, will feature increased bending losses and complicate the coupler design. In contrast, the large resonator needs both the signal and idler photons to be resonant, requiring careful design but featuring a geometry which is more readily fabricated.

\textit{Single-resonant ring:}  
The spacing between adjacent resonances scales inversely with the length of a single round trip of the resonator. Accordingly, this design is based on a short resonator, singly resonant at the signal frequency, where, due to interference effects, the incident photon undergoes effective bandwidth compression before being up-converted to the idler wavelength. We present the behavior of such a device in Fig. \ref{fig:FrequencyCompression}a, exemplary for a sum-frequency process driven by a cw-laser in the Lithium Niobate on-insulator (LNOI) platform, achieving bandwidth compression of up to three orders of magnitude. To avoid destructive interference at the idler wavelength, the generated idler photon needs to be out-coupled from the resonator with, ideally, unit efficiency. This can be achieved, for instance, by generating the idler photon in a distinct mode and adapting the width of the output waveguide to selectively couple this mode \cite{Zhao:23}, or by employing a wavelength dependent out-coupler. 

\textit{Doubly-resonant ring:} In a large resonator, the conversion region includes multiple resonances at the signal and idler wavelengths. Among these, only those that satisfy conservation of energy with the pump laser (see Fig. (\ref{fig:DesignSketch}b), experience efficient up-conversion. Accordingly, designing the resonator to be doubly-resonant at the interacting signal and idler-frequencies will result in up-conversion only at single, select frequencies. We present such a design, exemplary a for ring-resonator realized in LNOI operated with a cw-laser in Fig. \ref{fig:FrequencyCompression}b, featuring bandwidth compression of three orders of magnitude.

\section{Discussion \& Outlook}
With this article, we aimed to present devices capable of achieving QFC alongside significant spectral bandwidth compression to interface quantum memories with flying qubits, a capability we consider critical and well-attainable, but which has so far not been explored. We explicitly point toward devices based on single-photon sum-frequency generation in integrated photonics circuits, capable of achieving both QFC over hundreds of nanometers and spectral bandwidth compression by up to three orders of magnitude.

These devices are, in principle, based on the underlying physics, able to operate near-unit efficiency, but will perform below this ideal due to waveguide losses and fabrication errors associated with the couplers and the nonlinear waveguide. Sub-optimal coupling into the resonator, whether due to fabrication errors or constraints imposed by a small resonator, may further involve narrow coupling bandwidths, which could limit the range of available sum-frequency processes beyond the one considered her.

We note, that careful design of the phase-matching process described in Fig. \ref{fig:FrequencyConversion}, may readily be used to achieve the reverse process, connecting the narrowband memory regime to the broadband telecom regime.

We note further, that the high precision required in the fabrication of the resonator and the poling can be notably relaxed by adjusting the wavelength and bandwidth of the pump laser. This adjustment can ease constraints on the size of the single-resonant design \cite{AllgaierSilberhorn:17} or help align the double-resonance condition.

Alternatively, frequency conversion and bandwidth compression in the single-resonant design can be performed sequentially, with the nonlinear conversion occurring in a straight waveguide outside the resonator. The double-resonant design may be implemented using bulk-optics resonators, similar to approaches involving cavity-enhanced spontaneous parametric down-conversion \cite{SlatteryTang:19}.

Lastly, we would like to briefly point toward potential implementations of our device on emerging photonics platforms. Although a demonstration of QFC has not yet been shown in these platforms, etch-less strip-loaded waveguides \cite{YeSun:22,WangHu:17} promise to significantly simplify fabrication procedures, and plasmonic waveguides \cite{ThomaschewskiBozhevolnyi:22,SnowSiampour:24} allow for ultra-compact, although lossy, ring resonators.

This work contributes to ongoing efforts in quantum communication by providing generalized designs to more effectively connect different spectral-temporal regimes together with considerations concerning their experimental realization. Our findings support the development of more efficient and reliable long-range quantum information transfer, laying the groundwork for future research in integrated quantum networks.

\section*{Acknowledgments}
AP acknowledges an RMIT University Vice-Chancellor’s Senior Research Fellowship and a Google Faculty Research Award. This work was supported by the Australian Government through the Australian Research Council under the Centre of Excellence scheme (No: CE170100012).

\section*{Data Availability Statement}
All data generated or analyzed during this study are included in this published article.

{\appendix[] Fig. \ref{fig:AppendixA} depicts the waveguide geometry used for the calculations presented in Figs. \ref{fig:FrequencyConversion} and \ref{fig:FrequencyCompression}, together with the corresponding dispersion curves. This design was adopted from a state-of-the-art nonlinear integrated ring resonator \cite{LuTang:19}, but we note that the choice of waveguide geometry is largely arbitrary and, within reason, not essential to the device's functionality. Similarly, the explicit values of the pump power and bandwidth used in the simulation in the inset of Fig. \ref{fig:FrequencyConversion} may differ significantly from those of an eventual experimental implementation, both due to the available freedom of design and the considerable discrepancy between simulation and experiments observed in integrated (quantum) three-wave mixing \cite{ChenFan:24}.
\begin{figure}
    \centering
    \includegraphics[width=1\columnwidth]{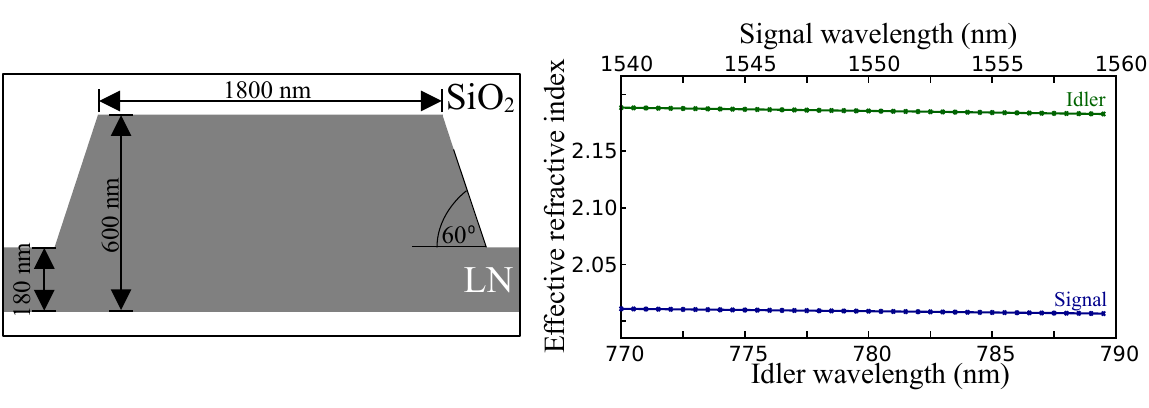}
    \caption{Schematic of the waveguide geometry used to calculate the results in section II, largely adapted from \cite{LuTang:19}, featuring a ridge-waveguide in a z-cut MgO doped $\mathrm{LiNbO_{3}}$ crystal, cladded with a layer of $\mathrm{SiO_{2}}$. The refractive indices used to calculate the dispersion were adapted from \cite{Zelmon:97}.}
    \label{fig:AppendixA}
\end{figure}
}

\bibliographystyle{IEEEtran}

\vfill

\end{document}